\documentclass[prl,twocolumn,superscriptaddress,floatfix,noshowpacs,10pt]{revtex4-1}%
\usepackage{graphicx,bm,times}
\usepackage{amsmath}
\usepackage{amsfonts}
\usepackage{amssymb}

\begin{document}

\title{Suppression of orbital ordering by chemical pressure in FeSe$_{1-x}$S$_x$}

\author{M. D. Watson}
\thanks{current affiliation: Diamond Light Source, Harwell Campus, Didcot, OX11 0DE, UK}
\affiliation{Clarendon Laboratory, Department of Physics,
University of Oxford, Parks Road, Oxford OX1 3PU, UK}
 
\author{T. K. Kim}
\affiliation{Diamond Light Source, Harwell Campus, Didcot, OX11 0DE, UK}

\author{A. A. Haghighirad}
\affiliation{Clarendon Laboratory, Department of Physics,
University of Oxford, Parks Road, Oxford OX1 3PU, UK}

\author{S. F. Blake}
\affiliation{Clarendon Laboratory, Department of Physics,
	University of Oxford, Parks Road, Oxford OX1 3PU, UK}

\author{N.~R.~Davies}
\affiliation{Clarendon Laboratory, Department of Physics,
	University of Oxford, Parks Road, Oxford OX1 3PU, UK}

\author{M.~Hoesch}
\affiliation{Diamond Light Source, Harwell Campus, Didcot, OX11 0DE, UK}


\author{T. Wolf}
\affiliation{Institute for Solid State Physics, Karlsruhe Institute of Technology, 76131 Karlsruhe, Germany}

\author{A. I. Coldea}
\email[corresponding author:~]{amalia.coldea@physics.ox.ac.uk}
\affiliation{Clarendon Laboratory, Department of Physics,
University of Oxford, Parks Road, Oxford OX1 3PU, UK}

\begin{abstract}

We report a high-resolution angle-resolved photo-emission spectroscopy study of the evolution of the electronic structure of FeSe$_{1-x}$S$_x$ single crystals. Isovalent S substitution onto the Se site constitutes a chemical pressure which subtly modifies the electronic structure of FeSe at high temperatures and induces a suppression of the tetragonal-symmetry-breaking structural transition temperature from 87~K to 58~K for $x=0.15$. With increasing S substitution, we find smaller splitting between bands with $d_{yz}$ and $d_{xz}$ orbital character and weaker anisotropic distortions of the low temperature Fermi surfaces. These effects evolve systematically as a function of both S substitution and temperature, providing strong evidence that an orbital ordering is the underlying order parameter of the structural transition in FeSe$_{1-x}$S$_x$. Finally, we detect the small inner hole pocket for $x$=0.12, which is pushed below the Fermi level in the orbitally-ordered low temperature Fermi surface of FeSe.
\end{abstract}
\date{\today}
\maketitle

The nature and origin of the ordered phases found in proximity to high temperature superconductivity relate directly to the mechanism of unconventional superconductivity \cite{Fernandes2014}. In the iron-based superconductors, the parent non-superconducting phases typically exhibit a magnetic stripe ordering, which breaks four-fold symmetry. However in certain regions of the phase diagram, the four-fold symmetry of the lattice is broken at a higher temperature $T_s$ than the onset of magnetism. One explanation is that this so-called {\it nematic} phase nevertheless originates from magnetic interactions, which in some models may pick out a particular direction to break fourfold symmetry without long-range magnetic ordering \cite{Fernandes2014}. Alternatively, it has been suggested that $T_s$ corresponds to an orbital instability breaking the degeneracy of the Fe $d_{xz}$ and $d_{yz}$ orbitals, distinct from magnetic order~\cite{Bohmer2014,Kontani2011}. FeSe provides a fascinating test case for these ideas, since it undergoes a structural transition at $T_s$=87~K but does not magnetically order~\cite{McQueen2009a} and has been the subject of intense theoretical study \cite{Mukherjee2015,Chubukov2015,Glasbrenner2015}. Detailed studies of the electronic structure in the tetragonal-symmetry-broken phase~\cite{Watson2015,Zhang2015,Shimojima2014,Zhang2015b,Suzuki2015} have found a prominent splitting of bands with $d_{xz}$ and $d_{yz}$ orbital character, which also manifests in strong distortions of the Fermi surface. However, this {\it orbital ordering} has been found to be sensitive to both physical pressure \cite{Terashima2015,Knoner2015} and substrate-induced strain in multilayer thin film samples \cite{Tan2013,Zhang2015b}. Chemical pressure, as in isovalently substituted FeSe$_{1-x}$S$_x$, provides another tuning parameter which can be used to study how the electronic structure evolves as orbital order is suppressed, helping to establish the nature of the order parameter associated with $T_s$, and determining to what extent the orbital fluctuations found in proximity to the ordered phase may play a role in the unconventional superconductivity.

In this paper we report a high-resolution angle-resolved photo-emission spectroscopy (ARPES) study of the electronic structure of S-substituted $\beta$-FeSe single crystals. In the high-temperature tetragonal phase, we find that the sizes of the Fermi surfaces and the Fermi velocities both increase with S substitution. The suppression of the structural transition $T_s$ by S substitution correlates with a reduction of the electronic anisotropy and the splitting of bands with $d_{xz}$ and $d_{yz}$ orbital characters, which onset at $T_s$. This provides strong evidence that the instability towards orbital ordering drives the fourfold symmetry-breaking transition. Finally, we show that the inner hole pocket which is pushed below the Fermi level in the orbitally-ordered low temperature Fermi surface of FeSe \cite{Watson2015} does cross the Fermi level in FeSe$_{1-x}$S$_x$ for $x$=0.12.

\begin{figure}
\centering
\includegraphics[width=0.7\linewidth]{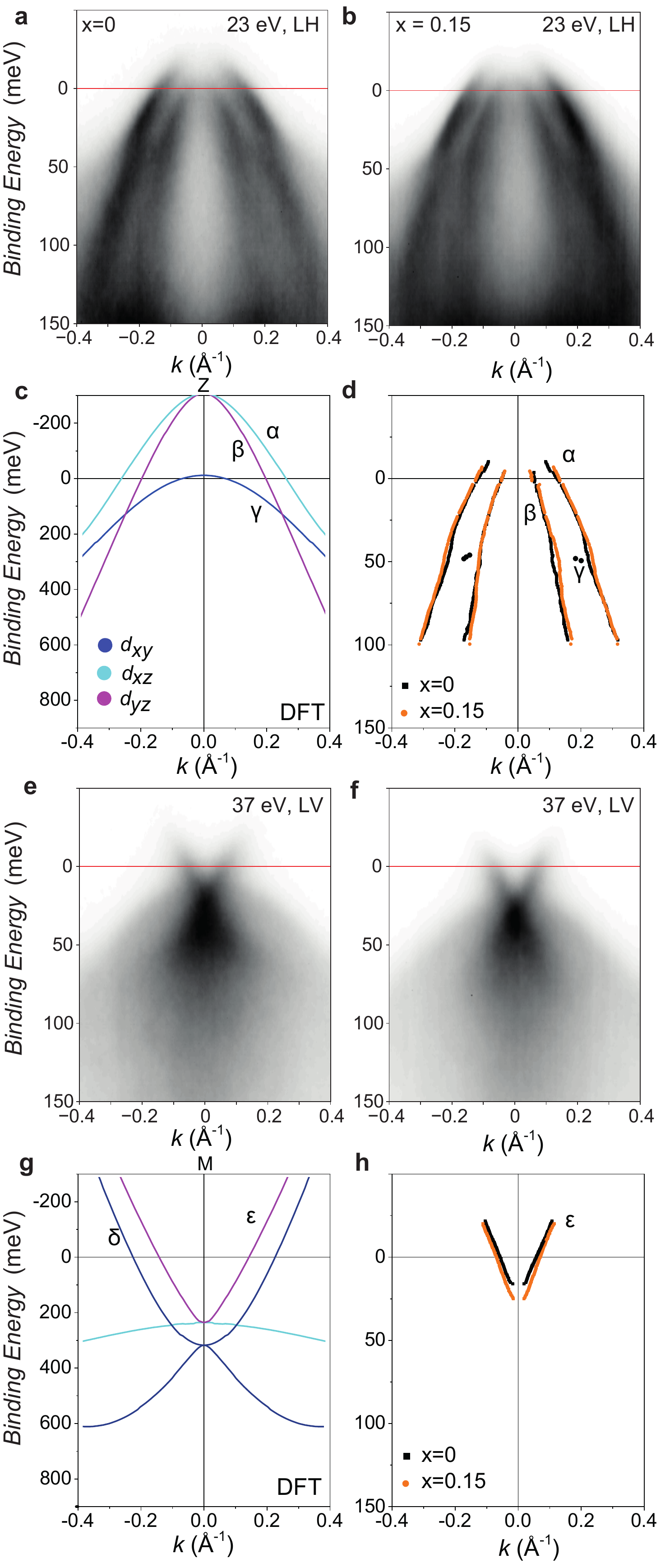}
\caption{(color online) {\bf High temperature electronic structure of FeSe$_{1-x}$S$_x$}.
 ARPES data for a) $x$=0 and b) $x$=0.15 for the hole-like bands at the Z point (measured using a photon energy of 23~eV). 
 c) Band dispersions from a DFT calculation at the Z point, compared to d) the extracted peak positions (from fits to MDCs) for $x$=0 and $x$=0.15. e-h) ARPES data, DFT calculation and extracted band positions for the electron pockets at the M point (at 37~eV).  The predicted outer electron band $\delta$ with $d_{xy}$ orbital character is not observed. Bands are labelled $\alpha-\epsilon$, as in previous studies \cite{Watson2015,Maletz2014}.}
\label{fig:Fig1}
\end{figure}

{\it Experimental Details.}
Samples were grown by the KCl/AlCl$_3$ chemical vapour transport method \cite{Chareev2013,Bohmer2013}.
ARPES  measurements  were  performed  at  the  I05  beamline  at the  Diamond
Light Source, UK. Single crystal samples with sizes $\approx~1\times1~$mm were cleaved in-situ below 15~K at a pressure
lower than 2$\times$10$^{-10}$ mbar. Samples were carefully oriented with the scattering plane along M-$\rm \Gamma$-M (Fe-Fe bond direction).
Measurements  were  performed  using  linearly-polarised  synchrotron
light at 20-80~eV, employing a Scienta R4000 hemispherical electron
energy  analyser. Band  structure  calculations  were  performed in Wien2k \cite{Wien2k} using the GGA approximation~\footnote{Since FeSe does not show magnetic order ARPES spectra are compared to non-spin-polarized band structure, with the relaxed lattice parameters $a = 3.7651$ \AA, $c$ = 5.5178 \AA, $z_{Se}$ = 0.24128 from Ref.~\onlinecite{Kumar2012}}.

{\it High temperature ARPES data.}
Fig.~\ref{fig:Fig1} shows a comparison of ARPES data of FeSe$_{1-x}$S$_x$ single crystals for $x$=0 and $x$=0.15 at temperatures above $T_s$. The quality of the ARPES spectra of the substituted crystals remains high, since the Fe $3d$ orbitals which dominate the low-energy electronic structure are not directly affected by the substitutional disorder outside of the Fe plane.  The comparison between the ARPES spectra and the density functional theory calculations (DFT) in Fig.~\ref{fig:Fig1} 
reveals several important differences, as has been previously reported~\cite{Watson2015,Zhang2015,Shimojima2014,Zhang2015b}. 
Firstly, all observed pockets are substantially smaller than DFT predictions~\cite{Watson2015}.
 Secondly, the predicted $\gamma$ hole band with $d_{xy}$ character (Fig.~\ref{fig:Fig1}(c)) is found to not cross the Fermi level but remains $\sim$50~meV below it (Fig.~\ref{fig:Fig1}(d)) with a particularly large band renormalisation of $\sim$8 compared to the other bands \cite{Watson2015,Maletz2014}. Thirdly, the outer  
 electron band $\delta$ with $d_{xy}$ character is not observed in experiments (Fig.~\ref{fig:Fig1}(g)).

The effect of substituting smaller S ions onto the Se site results in a lattice contraction \cite{Mizuguchi2009} and intuitively might result in a greater orbital overlap, increasing the bandwidth. 
In order to perform a quantitative comparison, we extract the band positions from constrained fits to the momentum distribution curves (MDCs), focusing on the high-symmetry cuts through the Z and M points, as shown in Fig.~\ref{fig:Fig1}(d,h) 
for $x$=0 and $x$=0.15 and summarized in Table~I in the Supplementary Material (SM). We
 find two systematic features as a function of S substitution; the Fermi surfaces generally 
 increase in size, and additionally the Fermi velocities ($dE/dk$ at $E_F$) increase, although both only by $\sim$10\%. 
These two trends may explain the suppression of $T_s$, since an increase in the size of Fermi surfaces is likely to reduce the degree of particle-hole Fermi surface nesting (which would be perfect in the limit of point-like Fermi surfaces), while the increase in Fermi velocities corresponds to a reduction in the density of states at the Fermi level, reducing the bare susceptibility and thereby suppressing the orbital instability. 

\begin{figure*}
\centering
\includegraphics[width=\linewidth]{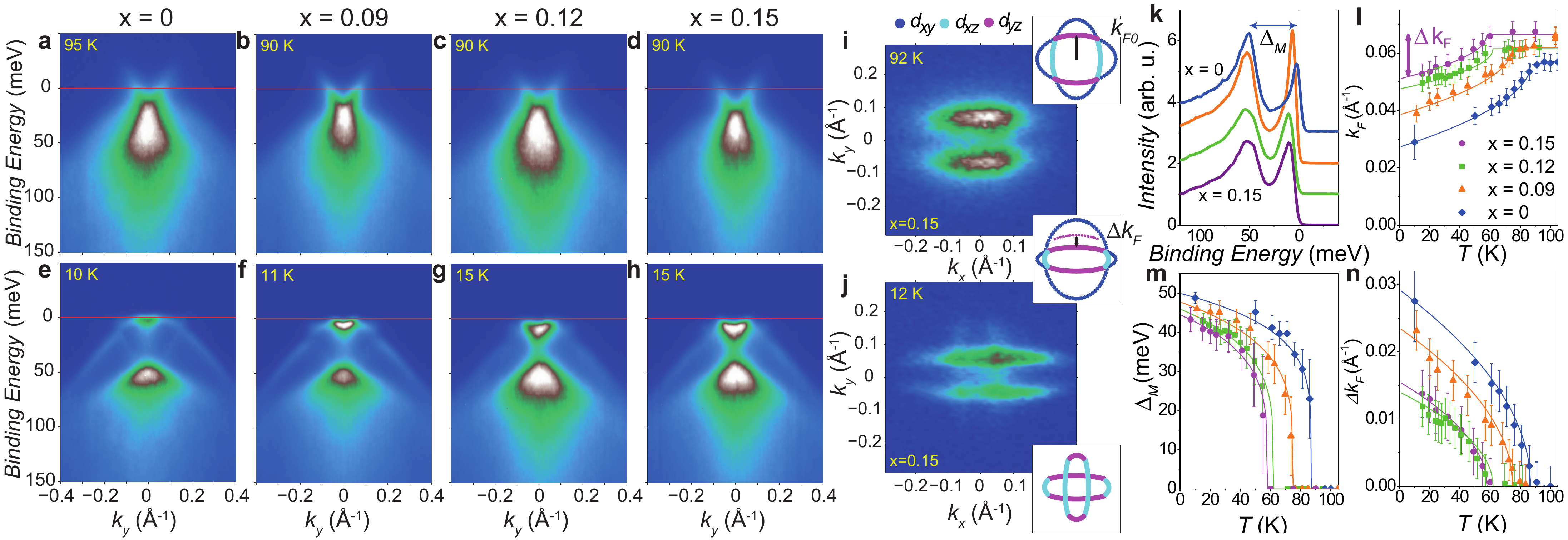}
\caption{(color online) {\bf Suppression of orbital ordering with S substitution on the electron pockets}. a-h) ARPES data for a cut through the M point ($\Gamma$-M-$\Gamma$ direction) at low and high temperatures for FeSe$_{1-x}$S$_x$ with $x=0 - 0.15$ showing the prominent splitting of intensity below $T_s$. i) High and j) low-temperature Fermi surface maps for $x$=0.15. Cartoon insets show schematic Fermi surface above $T_s$ (top), which is distorted below $T_s$ (middle), and the experimental case of twinned crystals where the $d_{xy}$ electron band is not observed (bottom). k) Comparison of low temperature EDCs at the M point showing the reduction of the splitting parameter, $\Delta_M$, with S substitution. l) Temperature dependences of the Fermi $k_F$ vector of the $d_{yz}$ portion of the Fermi surface, indicating a distortion $\Delta{}k_F$ which onsets at $T_s$. The temperature dependence of the extracted order parameters of the orbital ordering, m) $\Delta_M$ and n)$\Delta{}k_F$, as a function of S substitution, as discussed in the main text. Solid lines are guides to the eye.}
\label{fig:Fig2}
\end{figure*}

{\it Suppression of orbital ordering with S substitution.}
In Fig.~\ref{fig:Fig2}(a-h) we show how the electron pockets observed in a cut through the M point evolve with S substitution, above and below $T_s$. At high temperatures, only the $\epsilon$ band (with $d_{xz}/d_{yz}$ orbital character and defined in Fig.\ref{fig:Fig1}) is observed~
\footnote{We have never observed the outer $d_{xy}$ electron band, even in the second Brillouin zone where the weak signal due to matrix elements effects might be expected to be larger}. 
A prominent feature of ARPES studies of FeSe in both bulk crystals and multilayer thin films is a splitting of intensity at the M point below $T_s$, which we denote $\Delta_M$.  
This splitting, reaching $\Delta_M = 49(2)$~meV for bulk FeSe, corresponds to a splitting of the bands with 
$d_{yz}$ and $d_{xz}$ orbital characters at the M point in the Brillouin zone \cite{Shimojima2014,Watson2015,Zhang2015,Zhang2015b},
 which are degenerate in the high-temperature fourfold-symmetric phase. The value of $\Delta_M$
is much larger than what would be expected from DFT calculations 
 simply taking into account the small orthorhombic distortion of the lattice \cite{Watson2015}, which was taken to be one indication that the structural transition at $T_s$ is primarily an electronic instability, and more specifically due to orbital ordering. Therefore, the splitting $\Delta_M$, extracted from the positions of peaks in the energy distribution curve (EDC) at the M point (Fig.~\ref{fig:Fig2}(k)), is one experimental proxy order parameter for the orbital ordering which can be studied as a function of temperature and S substitution. 

The orbital ordering, which can alternatively be described as the development of an orbital polarisation $\Delta{}n=n_{xz}-n_{yz}$, also manifests as a distortion of the Fermi surface. 
As shown in the low-temperature Fermi surface map in Fig.~\ref{fig:Fig2}(j)), sections of the electron pockets with $d_{yz}$ orbital character contract while $d_{xz}$ sections expand, leaving an elongated and anisotropic Fermi surface, which is observed to form a cross-shape due to effect of sample twinning. 
Experimentally, we can extract the Fermi $k_F$-vector of the $d_{yz}$ portion 
of the electron pocket from peak fits of the MDC at $E_F$ (see Fig.~\ref{fig:Fig2}(e-h))
 and follow the Fermi surface distortion as a function of temperature, as shown in Fig.~\ref{fig:Fig2}(l)). 
This provides another  proxy order parameter for the orbital ordering,
 which we define as $\Delta{}k_F=k_F-k_{F0}$ \cite{Watson2015}, where $k_{F0}$ is the Fermi $k$-vector just above $T_s$ in the tetragonal phase. 

Figs.~\ref{fig:Fig2}(m,n) show how these two proxy order parameters 
for the orbital ordering, 
the band splitting $\Delta{}_M$ and Fermi surface distortion $\Delta{}k_F$,
both evolve systematically as a function of temperature and S substitution. The onset of both parameters coincides with $T_s$ 
(as determined separately from anomalies in resistivity measurements \cite{FeSeS_QOs}) and both follow an order-parameter-like temperature dependence, as shown in Fig.~\ref{fig:Fig2}(m,n). 
The fact that both order parameters change systematically with S substitution 
is strong evidence that the reduction of orbital ordering is directly related to the suppression of $T_s$ in FeSe$_{1-x}$S$_x$
(see also Fig.~\ref{fig:Fig4}).

\begin{figure}
\centering
\includegraphics[width=0.85\linewidth]{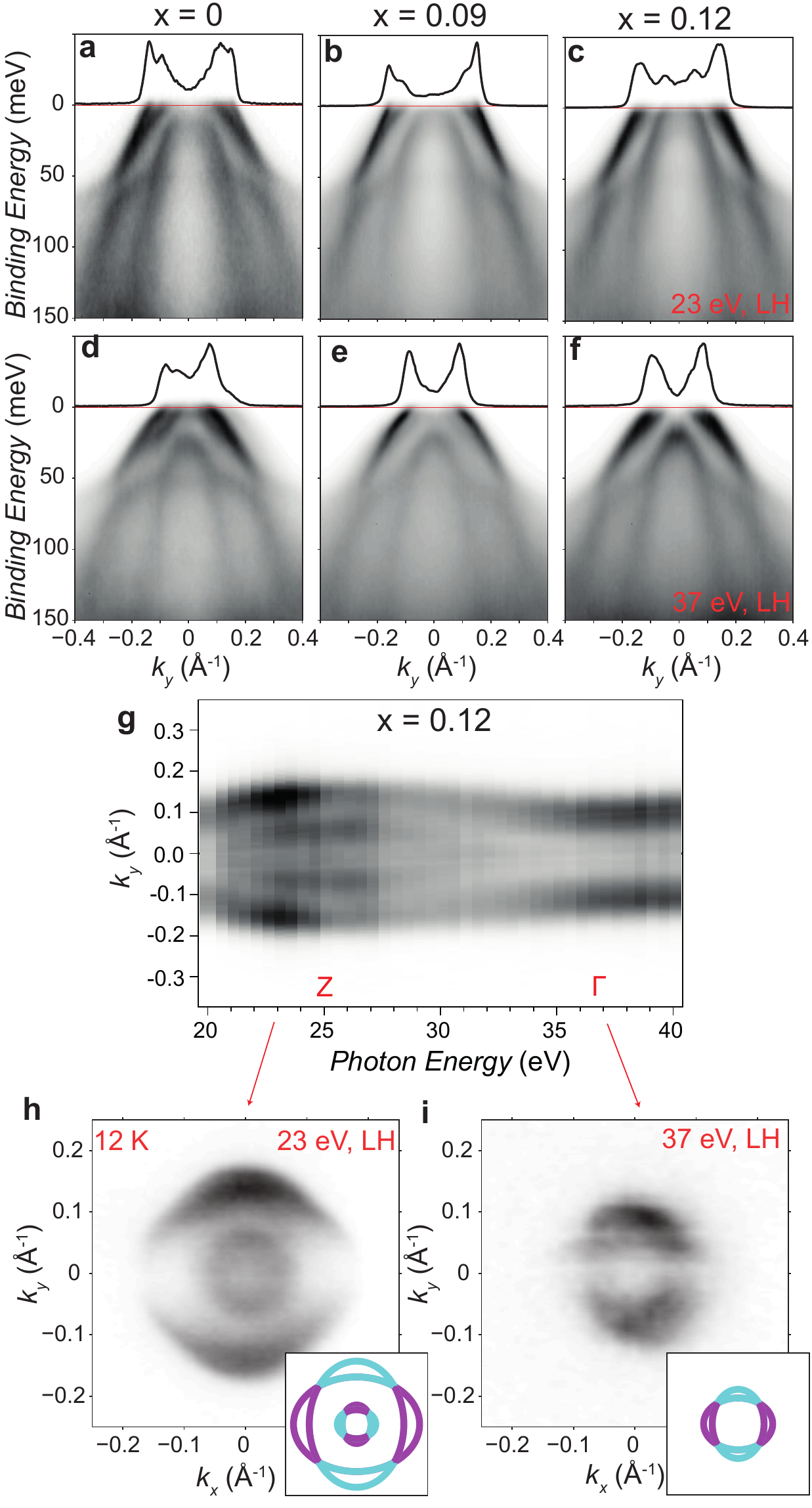}
\caption{(color online) {\bf Emergence of the small inner hole band at low temperatures for $x$=0.12}. a-c) ARPES data for a high-symmetry cut through the centre of the Brillouin zone at 23 eV (near the Z point) at $T\approx$11~K for $x$=0-0.12. Solid black lines are the MDCs at the Fermi level, indicating the splitting of the outer hole band $\alpha$
 and the emergence of the inner $\beta$ band for $x=0.12$. d-f) As above, but for a cut at the $\rm \Gamma$ point (37 eV). g) Photon energy-dependence of the Fermi level MDC at $x=0.12$, indicating that the inner hole band forms a 3D pocket around the Z point, as also shown by the Fermi surface maps at h) the $Z$ (23 eV) and i) $\Gamma$ point (37~eV). Cartoon insets show how the apparent splitting of bands arises from Fermi surface distortion and sample twinning.}
\label{fig:Fig3}
\end{figure}

{\it The emergence of the inner hole band for $x$=0.12 at low temperatures.}
We now consider the evolution of the hole pockets with S substitution at low temperatures, as shown in Fig.~\ref{fig:Fig3}. Similar to the electron pockets, the degree of Fermi surface anisotropy on the hole pockets reduces with S substitution, as indicated by the reduced splitting of the two $k_F$ values of the outer hole band, $\alpha$, as shown in Fig.~\ref{fig:Fig3}(a-f) and Table~I in SM. In all samples at high temperature the inner $\beta$ hole band forms a tiny 3D pocket around Z (see Fig.~\ref{fig:Fig1}(a,b)), where a splitting of $\sim$20~meV \cite{Watson2015} between the $\alpha$ and $\beta$ hole bands occurs due to spin-orbit coupling only \cite{Fernandes2014b}. At low temperature an extra splitting associated with the orbital ordering \cite{Fernandes2014b} pushes the $\beta$ band below the Fermi surface \cite{Watson2015,Suzuki2015} in FeSe, as can be seen by the lack of features in the Fermi level MDC for FeSe in Fig.~\ref{fig:Fig3}(a). However, as the orbital ordering is reduced with S substitution, this inner pocket can exist at low temperatures once more, and is clearly observed at $x~=~0.12$ (Fig.~\ref{fig:Fig3}(c)). The photon-energy dependence shown in Fig.~\ref{fig:Fig3}(g) (which is equivalent to a $k_z$ dispersion) shows that it forms a 3D pocket around Z, not extending throughout the Brillouin zone, and also seen in the detailed map of the Fermi surface around the Z point in Fig.~\ref{fig:Fig3}(h). This pocket will also break four-fold symmetry, as indicated by the inset to Fig.~\ref{fig:Fig3}(h). This new hole pocket could appear as an additional low frequency in quantum oscillations measurements \cite{FeSeS_QOs} in addition to those corresponding to three different bands in previous reports \cite{Watson2015,Watson2015b,Terashima2014}.

\begin{figure}
\centering
\includegraphics[width=0.8\linewidth]{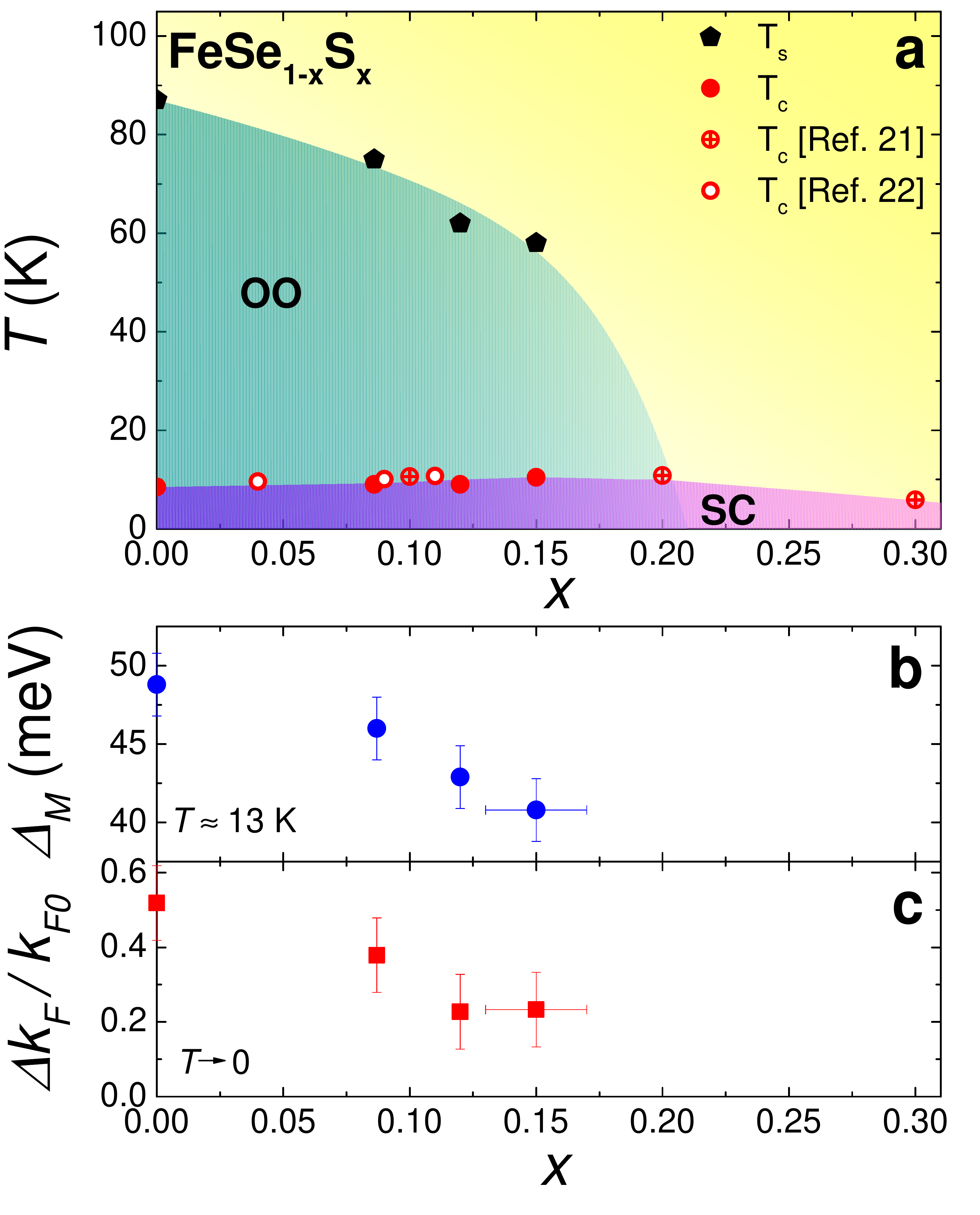}
\caption{a) {(color online)\bf Proposed phase diagram of FeSe$_{1-x}$S$_x$}. $T_s$ and $T_c$ values based on our resistivity measurements \cite{FeSeS_QOs}, 
compared to literature values of $T_c$ reported in Ref.~\cite{Mizuguchi2009} (resistivity midpoint) and Ref.~\cite{Abdel-Hafiez2015}
(heat capacity). OO and SC refer to orbital order and the superconducting phase respectively. 
b,c) Evolution of the band splitting, $\Delta_M$, and Fermi surface deformation, $\Delta{}k_F/k_{F0}$, with S substitution at low temperature, demonstrating that the magnitude of the orbital ordering is directly correlated with the suppression of $T_s$.}
\label{fig:Fig4}
\end{figure}

Fig.~\ref{fig:Fig4}(a) shows a proposed phase diagram of FeSe$_{1-x}$S$_x$
based on the experimental values of the structural and superconducting 
transition temperatures extracted from our resistivity measurements 
(to be presented elsewhere~\cite{FeSeS_QOs}) and Refs.~\cite{Mizuguchi2009,Abdel-Hafiez2015}.
With S substitution there is a strong suppression of the structural transition,
whereas the $T_c$ of the superconducting state remains relatively 
constant for low $x$. As shown in Fig.~\ref{fig:Fig4}(b,c), both the low temperature $\Delta_M$ and $\Delta{}k_F/k_{F0}$ extracted 
from our ARPES data reduce with $x$, correlating with $T_s$ in the phase diagram. 
A similar trend was also found in thin films \cite{Tan2013,Zhang2015}, 
where $\Delta_M$ and $T_s$ are found to increase for thinner samples which are more strained. Thus, the systematic change of the degree of orbital 
ordering with S substitution, as measured by our two experimental proxy order parameters, indicates that the true electronic order parameter associated with the structural transition in FeSe$_{1-x}$S$_x$ is the orbital polarisation $\Delta{}n=n_{xz}-n_{yz}$.
Therefore, the transition at $T_s$ corresponds to an instability of the electronic structure of this 
iron-based superconductor in the charge/orbital channel, which is distinct from the stripe spin density wave phase observed in other systems such as BaFe$_2$As$_2$.

Studies of FeSe under applied physical pressure have found that the structural transition $T_s$ is initially suppressed in a similar manner to the chemical pressure effect of S substitution \cite{Terashima2015,Knoner2015}, although the magnetic phase \cite{Terashima2015,Bendele2010} and high temperature superconductivity \cite{Medvedev2009} found under higher physical pressures have not been reported to occur in FeSe$_{1-x}$S$_x$ \cite{Mizuguchi2009,Abdel-Hafiez2015} (at ambient pressure \cite{Tomita2015}). NMR studies found that spin fluctuations in FeSe increase in magnitude with physical pressure \cite{Imai2009} while $T_s$ is suppressed \cite{Terashima2015,Knoner2015}, suggesting that magnetic interactions are not crucial for the structural instability.
On the other hand, our ARPES results show a direct correlation between the decrease in $T_s$ and the suppression of orbital ordering under chemical pressure (Fig.~\ref{fig:Fig4}). 
However, the fact that the superconductivity in FeSe$_{1-x}$S$_x$ seems to be almost independent of the orbital ordering and there is no pronounced superconducting dome in the absence of any magnetic phase (Fig.~\ref{fig:Fig4}(a)) counts against orbital fluctuations providing a major component of the unconventional pairing mechanism. The small size of all Fermi surfaces and the absence of the inner hole pocket in particular gives FeSe a very different electronic structure to other Fe-based superconductors. However the emergence of the inner hole pocket in FeSe$_{1-x}$S$_x$ could have a significant impact on the susceptibility and pairing interactions, and it is also likely to emerge when orbital ordering is suppressed under physical pressure in FeSe. The exact nature of the orbital ordering is likely to be more complex than simply a uniform site-centred ferro-orbital ordering between $d_{xz}$ and $d_{yz}$ orbitals \cite{Mukherjee2015,Fernandes2014b,Jiang2015,Suzuki2015}, and future studies using strain to detwin samples may shed light on the details of the orbital mechanism. An alternative viewpoint is that the Fermi surface deformations which are suppressed with S substitution could be interpreted as a manifestation of a d-wave Pomeranchuk instability, which is predicted to be weakened as the size of Fermi surface increases\cite{ChubukovPC}. 

In summary, our high-resolution ARPES study of FeSe$_{1-x}$S$_x$ single crystals has established that the chemical pressure increases the size of the Fermi surface pockets and Fermi velocities at high temperature. Most importantly, we have shown the decrease of $T_s$ is associated with a suppression of orbital order  with S substitution, by observing a reduced splitting of $d_{xz}$ and $d_{yz}$ orbitals and weaker Fermi surface anisotropy. These results demonstrate that this orbital ordering is the true order parameter of the tetragonal-symmetry-breaking transition at $T_s$ in FeSe. Finally, these subtle changes to the electronic structure lead to the appearance of the inner hole band in the low temperature Fermi surface. FeSe$_{1-x}$S$_x$ is emerging as a key system to help resolve long-standing controversies related to role of spin and orbital degrees of freedom in the iron-based superconductors.

\begin{acknowledgments}
We acknowledge fruitful discussions with T.~Scaffidi, A.~Schofield, P. Hirschfeld, O.~Vafek, A.~Chubukov and C.~Meingast. We thank A.~Narayanan, P.~Schoenherr, L.~J.~Collins-Macintyre and T.~Hesjedal for technical support.
This work was mainly supported by the EPSRC
(EP/L001772/1, EP/I004475/1, EP/I017836/1). We thank
Diamond Light Source for access to Beamline I05 (proposal
number SI11792) that contributed to the results presented
here. The authors would like to acknowledge the use of the
University of Oxford Advanced Research Computing (ARC)
facility in carrying out part of this work, http://dx.doi.org/10.5281/zenodo.22558. A.I.C. acknowledges
an EPSRC Career Acceleration Fellowship (EP/I004475/1).
\end{acknowledgments}


\begin{thebibliography}{34}%
\makeatletter
\providecommand \@ifxundefined [1]{%
 \@ifx{#1\undefined}
}%
\providecommand \@ifnum [1]{%
 \ifnum #1\expandafter \@firstoftwo
 \else \expandafter \@secondoftwo
 \fi
}%
\providecommand \@ifx [1]{%
 \ifx #1\expandafter \@firstoftwo
 \else \expandafter \@secondoftwo
 \fi
}%
\providecommand \natexlab [1]{#1}%
\providecommand \enquote  [1]{``#1''}%
\providecommand \bibnamefont  [1]{#1}%
\providecommand \bibfnamefont [1]{#1}%
\providecommand \citenamefont [1]{#1}%
\providecommand \href@noop [0]{\@secondoftwo}%
\providecommand \href [0]{\begingroup \@sanitize@url \@href}%
\providecommand \@href[1]{\@@startlink{#1}\@@href}%
\providecommand \@@href[1]{\endgroup#1\@@endlink}%
\providecommand \@sanitize@url [0]{\catcode `\\12\catcode `\$12\catcode
  `\&12\catcode `\#12\catcode `\^12\catcode `\_12\catcode `\%12\relax}%
\providecommand \@@startlink[1]{}%
\providecommand \@@endlink[0]{}%
\providecommand \url  [0]{\begingroup\@sanitize@url \@url }%
\providecommand \@url [1]{\endgroup\@href {#1}{\urlprefix }}%
\providecommand \urlprefix  [0]{URL }%
\providecommand \Eprint [0]{\href }%
\providecommand \doibase [0]{http://dx.doi.org/}%
\providecommand \selectlanguage [0]{\@gobble}%
\providecommand \bibinfo  [0]{\@secondoftwo}%
\providecommand \bibfield  [0]{\@secondoftwo}%
\providecommand \translation [1]{[#1]}%
\providecommand \BibitemOpen [0]{}%
\providecommand \bibitemStop [0]{}%
\providecommand \bibitemNoStop [0]{.\EOS\space}%
\providecommand \EOS [0]{\spacefactor3000\relax}%
\providecommand \BibitemShut  [1]{\csname bibitem#1\endcsname}%
\let\auto@bib@innerbib\@empty
\bibitem [{\citenamefont {Fernandes}\ \emph {et~al.}(2014)\citenamefont
  {Fernandes}, \citenamefont {Chubukov},\ and\ \citenamefont
  {Schmalian}}]{Fernandes2014}%
  \BibitemOpen
  \bibfield  {author} {\bibinfo {author} {\bibfnamefont {R.~M.}\ \bibnamefont
  {Fernandes}}, \bibinfo {author} {\bibfnamefont {A.~V.}\ \bibnamefont
  {Chubukov}}, \ and\ \bibinfo {author} {\bibfnamefont {J.}~\bibnamefont
  {Schmalian}},\ }\href {\doibase 10.1038/NPHYS2877} {\bibfield  {journal}
  {\bibinfo  {journal} {Nat. Phys.}\ }\textbf {\bibinfo {volume} {10}},\
  \bibinfo {pages} {97} (\bibinfo {year} {2014})}\BibitemShut {NoStop}%
\bibitem [{\citenamefont {B\"{o}hmer}\ \emph {et~al.}(2015)\citenamefont
  {B\"{o}hmer}, \citenamefont {Arai}, \citenamefont {Hardy}, \citenamefont
  {Hattori}, \citenamefont {Iye}, \citenamefont {Wolf}, \citenamefont {von
  L\"{o}hneysen}, \citenamefont {Ishida},\ and\ \citenamefont
  {Meingast}}]{Bohmer2014}%
  \BibitemOpen
  \bibfield  {author} {\bibinfo {author} {\bibfnamefont {A.~E.}\ \bibnamefont
  {B\"{o}hmer}}, \bibinfo {author} {\bibfnamefont {T.}~\bibnamefont {Arai}},
  \bibinfo {author} {\bibfnamefont {F.}~\bibnamefont {Hardy}}, \bibinfo
  {author} {\bibfnamefont {T.}~\bibnamefont {Hattori}}, \bibinfo {author}
  {\bibfnamefont {T.}~\bibnamefont {Iye}}, \bibinfo {author} {\bibfnamefont
  {T.}~\bibnamefont {Wolf}}, \bibinfo {author} {\bibfnamefont {H.~V.}\
  \bibnamefont {von L\"{o}hneysen}}, \bibinfo {author} {\bibfnamefont
  {K.}~\bibnamefont {Ishida}}, \ and\ \bibinfo {author} {\bibfnamefont
  {C.}~\bibnamefont {Meingast}},\ }\href {\doibase
  10.1103/PhysRevLett.114.027001} {\bibfield  {journal} {\bibinfo  {journal}
  {Phys. Rev. Lett.}\ }\textbf {\bibinfo {volume} {114}},\ \bibinfo {pages}
  {27001} (\bibinfo {year} {2015})}\BibitemShut {NoStop}%
\bibitem [{\citenamefont {Kontani}\ \emph {et~al.}(2011)\citenamefont
  {Kontani}, \citenamefont {Saito},\ and\ \citenamefont {Onari}}]{Kontani2011}%
  \BibitemOpen
  \bibfield  {author} {\bibinfo {author} {\bibfnamefont {H.}~\bibnamefont
  {Kontani}}, \bibinfo {author} {\bibfnamefont {T.}~\bibnamefont {Saito}}, \
  and\ \bibinfo {author} {\bibfnamefont {S.}~\bibnamefont {Onari}},\ }\href
  {\doibase 10.1103/PhysRevB.84.024528} {\bibfield  {journal} {\bibinfo
  {journal} {Phys. Rev. B}\ }\textbf {\bibinfo {volume} {84}},\ \bibinfo
  {pages} {024528} (\bibinfo {year} {2011})}\BibitemShut {NoStop}%
\bibitem [{\citenamefont {McQueen}\ \emph {et~al.}(2009)\citenamefont
  {McQueen}, \citenamefont {Williams}, \citenamefont {Stephens}, \citenamefont
  {Tao}, \citenamefont {Zhu}, \citenamefont {Ksenofontov}, \citenamefont
  {Casper}, \citenamefont {Felser},\ and\ \citenamefont {Cava}}]{McQueen2009a}%
  \BibitemOpen
  \bibfield  {author} {\bibinfo {author} {\bibfnamefont {T.~M.}\ \bibnamefont
  {McQueen}}, \bibinfo {author} {\bibfnamefont {A.~J.}\ \bibnamefont
  {Williams}}, \bibinfo {author} {\bibfnamefont {P.~W.}\ \bibnamefont
  {Stephens}}, \bibinfo {author} {\bibfnamefont {J.}~\bibnamefont {Tao}},
  \bibinfo {author} {\bibfnamefont {Y.}~\bibnamefont {Zhu}}, \bibinfo {author}
  {\bibfnamefont {V.}~\bibnamefont {Ksenofontov}}, \bibinfo {author}
  {\bibfnamefont {F.}~\bibnamefont {Casper}}, \bibinfo {author} {\bibfnamefont
  {C.}~\bibnamefont {Felser}}, \ and\ \bibinfo {author} {\bibfnamefont {R.~J.}\
  \bibnamefont {Cava}},\ }\href {\doibase 10.1103/PhysRevLett.103.057002}
  {\bibfield  {journal} {\bibinfo  {journal} {Phys. Rev. Lett.}\ }\textbf
  {\bibinfo {volume} {103}},\ \bibinfo {pages} {057002} (\bibinfo {year}
  {2009})}\BibitemShut {NoStop}%
\bibitem [{\citenamefont {Mukherjee}\ \emph {et~al.}(2015)\citenamefont
  {Mukherjee}, \citenamefont {Kreisel}, \citenamefont {Hirschfeld},\ and\
  \citenamefont {Andersen}}]{Mukherjee2015}%
  \BibitemOpen
  \bibfield  {author} {\bibinfo {author} {\bibfnamefont {S.}~\bibnamefont
  {Mukherjee}}, \bibinfo {author} {\bibfnamefont {A.}~\bibnamefont {Kreisel}},
  \bibinfo {author} {\bibfnamefont {P.~J.}\ \bibnamefont {Hirschfeld}}, \ and\
  \bibinfo {author} {\bibfnamefont {B.~M.}\ \bibnamefont {Andersen}},\ }\href
  {\doibase 10.1103/PhysRevLett.115.026402} {\bibfield  {journal} {\bibinfo
  {journal} {Phys. Rev. Lett.}\ }\textbf {\bibinfo {volume} {115}},\ \bibinfo
  {pages} {026402} (\bibinfo {year} {2015})}\BibitemShut {NoStop}%
\bibitem [{\citenamefont {Chubukov}\ \emph {et~al.}(2015)\citenamefont
  {Chubukov}, \citenamefont {Fernandes},\ and\ \citenamefont
  {Schmalian}}]{Chubukov2015}%
  \BibitemOpen
  \bibfield  {author} {\bibinfo {author} {\bibfnamefont {A.~V.}\ \bibnamefont
  {Chubukov}}, \bibinfo {author} {\bibfnamefont {R.~M.}\ \bibnamefont
  {Fernandes}}, \ and\ \bibinfo {author} {\bibfnamefont {J.}~\bibnamefont
  {Schmalian}},\ }\href {\doibase 10.1103/PhysRevB.91.201105} {\bibfield
  {journal} {\bibinfo  {journal} {Phys. Rev. B}\ }\textbf {\bibinfo {volume}
  {91}},\ \bibinfo {pages} {201105} (\bibinfo {year} {2015})}\BibitemShut
  {NoStop}%
\bibitem [{\citenamefont {Glasbrenner}\ \emph {et~al.}(2015)\citenamefont
  {Glasbrenner}, \citenamefont {Mazin}, \citenamefont {Jeschke},\ and\
  \citenamefont {Hirschfeld}}]{Glasbrenner2015}%
  \BibitemOpen
  \bibfield  {author} {\bibinfo {author} {\bibfnamefont {J.~K.}\ \bibnamefont
  {Glasbrenner}}, \bibinfo {author} {\bibfnamefont {I.~I.}\ \bibnamefont
  {Mazin}}, \bibinfo {author} {\bibfnamefont {H.~O.}\ \bibnamefont {Jeschke}},
  \ and\ \bibinfo {author} {\bibfnamefont {P.~J.}\ \bibnamefont {Hirschfeld}},\
  }\href@noop {} {\bibfield  {journal} {\bibinfo  {journal} {arXiv:1501.04946}\
  } (\bibinfo {year} {2015})}\BibitemShut {NoStop}%
\bibitem [{\citenamefont {Watson}\ \emph {et~al.}(2015)\citenamefont {Watson},
  \citenamefont {Kim}, \citenamefont {Haghighirad}, \citenamefont {Davies},
  \citenamefont {McCollam}, \citenamefont {Narayanan}, \citenamefont {Blake},
  \citenamefont {Chen}, \citenamefont {Ghannadzadeh}, \citenamefont
  {Schofield}, \citenamefont {Hoesch}, \citenamefont {Meingast}, \citenamefont
  {Wolf},\ and\ \citenamefont {Coldea}}]{Watson2015}%
  \BibitemOpen
  \bibfield  {author} {\bibinfo {author} {\bibfnamefont {M.~D.}\ \bibnamefont
  {Watson}}, \bibinfo {author} {\bibfnamefont {T.~K.}\ \bibnamefont {Kim}},
  \bibinfo {author} {\bibfnamefont {A.~A.}\ \bibnamefont {Haghighirad}},
  \bibinfo {author} {\bibfnamefont {N.~R.}\ \bibnamefont {Davies}}, \bibinfo
  {author} {\bibfnamefont {A.}~\bibnamefont {McCollam}}, \bibinfo {author}
  {\bibfnamefont {A.}~\bibnamefont {Narayanan}}, \bibinfo {author}
  {\bibfnamefont {S.~F.}\ \bibnamefont {Blake}}, \bibinfo {author}
  {\bibfnamefont {Y.~L.}\ \bibnamefont {Chen}}, \bibinfo {author}
  {\bibfnamefont {S.}~\bibnamefont {Ghannadzadeh}}, \bibinfo {author}
  {\bibfnamefont {A.~J.}\ \bibnamefont {Schofield}}, \bibinfo {author}
  {\bibfnamefont {M.}~\bibnamefont {Hoesch}}, \bibinfo {author} {\bibfnamefont
  {C.}~\bibnamefont {Meingast}}, \bibinfo {author} {\bibfnamefont
  {T.}~\bibnamefont {Wolf}}, \ and\ \bibinfo {author} {\bibfnamefont {A.~I.}\
  \bibnamefont {Coldea}},\ }\href {\doibase 10.1103/PhysRevB.91.155106}
  {\bibfield  {journal} {\bibinfo  {journal} {Phys. Rev. B}\ }\textbf {\bibinfo
  {volume} {91}},\ \bibinfo {pages} {155106} (\bibinfo {year}
  {2015})}\BibitemShut {NoStop}%
\bibitem [{\citenamefont {Zhang}\ \emph
  {et~al.}(2015{\natexlab{a}})\citenamefont {Zhang}, \citenamefont {Qian},
  \citenamefont {Richard}, \citenamefont {Wang}, \citenamefont {Miao},
  \citenamefont {Lv}, \citenamefont {Fu}, \citenamefont {Wolf}, \citenamefont
  {Meingast}, \citenamefont {Wu}, \citenamefont {Wang}, \citenamefont {Hu},\
  and\ \citenamefont {Ding}}]{Zhang2015}%
  \BibitemOpen
  \bibfield  {author} {\bibinfo {author} {\bibfnamefont {P.}~\bibnamefont
  {Zhang}}, \bibinfo {author} {\bibfnamefont {T.}~\bibnamefont {Qian}},
  \bibinfo {author} {\bibfnamefont {P.}~\bibnamefont {Richard}}, \bibinfo
  {author} {\bibfnamefont {X.~P.}\ \bibnamefont {Wang}}, \bibinfo {author}
  {\bibfnamefont {H.}~\bibnamefont {Miao}}, \bibinfo {author} {\bibfnamefont
  {B.~Q.}\ \bibnamefont {Lv}}, \bibinfo {author} {\bibfnamefont {B.~B.}\
  \bibnamefont {Fu}}, \bibinfo {author} {\bibfnamefont {T.}~\bibnamefont
  {Wolf}}, \bibinfo {author} {\bibfnamefont {C.}~\bibnamefont {Meingast}},
  \bibinfo {author} {\bibfnamefont {X.~X.}\ \bibnamefont {Wu}}, \bibinfo
  {author} {\bibfnamefont {Z.~Q.}\ \bibnamefont {Wang}}, \bibinfo {author}
  {\bibfnamefont {J.~P.}\ \bibnamefont {Hu}}, \ and\ \bibinfo {author}
  {\bibfnamefont {H.}~\bibnamefont {Ding}},\ }\href {\doibase
  10.1103/PhysRevB.91.214503} {\bibfield  {journal} {\bibinfo  {journal} {Phys.
  Rev. B}\ }\textbf {\bibinfo {volume} {91}},\ \bibinfo {pages} {214503}
  (\bibinfo {year} {2015}{\natexlab{a}})}\BibitemShut {NoStop}%
\bibitem [{\citenamefont {Shimojima}\ \emph {et~al.}(2014)\citenamefont
  {Shimojima}, \citenamefont {Suzuki}, \citenamefont {Sonobe}, \citenamefont
  {Nakamura}, \citenamefont {Sakano}, \citenamefont {Omachi}, \citenamefont
  {Yoshioka}, \citenamefont {Kuwata-Gonokami}, \citenamefont {Ono},
  \citenamefont {Kumigashira}, \citenamefont {B\"{o}hmer}, \citenamefont
  {Hardy}, \citenamefont {Wolf}, \citenamefont {Meingast}, \citenamefont
  {L\"{o}hneysen}, \citenamefont {Ikeda},\ and\ \citenamefont
  {Ishizaka}}]{Shimojima2014}%
  \BibitemOpen
  \bibfield  {author} {\bibinfo {author} {\bibfnamefont {T.}~\bibnamefont
  {Shimojima}}, \bibinfo {author} {\bibfnamefont {Y.}~\bibnamefont {Suzuki}},
  \bibinfo {author} {\bibfnamefont {T.}~\bibnamefont {Sonobe}}, \bibinfo
  {author} {\bibfnamefont {A.}~\bibnamefont {Nakamura}}, \bibinfo {author}
  {\bibfnamefont {M.}~\bibnamefont {Sakano}}, \bibinfo {author} {\bibfnamefont
  {J.}~\bibnamefont {Omachi}}, \bibinfo {author} {\bibfnamefont
  {K.}~\bibnamefont {Yoshioka}}, \bibinfo {author} {\bibfnamefont
  {M.}~\bibnamefont {Kuwata-Gonokami}}, \bibinfo {author} {\bibfnamefont
  {K.}~\bibnamefont {Ono}}, \bibinfo {author} {\bibfnamefont {H.}~\bibnamefont
  {Kumigashira}}, \bibinfo {author} {\bibfnamefont {A.~E.}\ \bibnamefont
  {B\"{o}hmer}}, \bibinfo {author} {\bibfnamefont {F.}~\bibnamefont {Hardy}},
  \bibinfo {author} {\bibfnamefont {T.}~\bibnamefont {Wolf}}, \bibinfo {author}
  {\bibfnamefont {C.}~\bibnamefont {Meingast}}, \bibinfo {author}
  {\bibfnamefont {H.~V.}\ \bibnamefont {L\"{o}hneysen}}, \bibinfo {author}
  {\bibfnamefont {H.}~\bibnamefont {Ikeda}}, \ and\ \bibinfo {author}
  {\bibfnamefont {K.}~\bibnamefont {Ishizaka}},\ }\href {\doibase
  10.1103/PhysRevB.90.121111} {\bibfield  {journal} {\bibinfo  {journal} {Phys.
  Rev. B}\ }\textbf {\bibinfo {volume} {90}},\ \bibinfo {pages} {121111}
  (\bibinfo {year} {2014})}\BibitemShut {NoStop}%
\bibitem [{\citenamefont {Zhang}\ \emph
  {et~al.}(2015{\natexlab{b}})\citenamefont {Zhang}, \citenamefont {Yi},
  \citenamefont {Liu}, \citenamefont {Li}, \citenamefont {Lee}, \citenamefont
  {Moore}, \citenamefont {Hashimoto}, \citenamefont {Masamichi}, \citenamefont
  {Eisaki}, \citenamefont {Mo}, \citenamefont {Hussain}, \citenamefont
  {Devereaux}, \citenamefont {Shen},\ and\ \citenamefont {Lu}}]{Zhang2015b}%
  \BibitemOpen
  \bibfield  {author} {\bibinfo {author} {\bibfnamefont {Y.}~\bibnamefont
  {Zhang}}, \bibinfo {author} {\bibfnamefont {M.}~\bibnamefont {Yi}}, \bibinfo
  {author} {\bibfnamefont {Z.-K.}\ \bibnamefont {Liu}}, \bibinfo {author}
  {\bibfnamefont {W.}~\bibnamefont {Li}}, \bibinfo {author} {\bibfnamefont
  {J.~J.}\ \bibnamefont {Lee}}, \bibinfo {author} {\bibfnamefont {R.~G.}\
  \bibnamefont {Moore}}, \bibinfo {author} {\bibfnamefont {M.}~\bibnamefont
  {Hashimoto}}, \bibinfo {author} {\bibfnamefont {N.}~\bibnamefont
  {Masamichi}}, \bibinfo {author} {\bibfnamefont {H.}~\bibnamefont {Eisaki}},
  \bibinfo {author} {\bibfnamefont {S.-K.}\ \bibnamefont {Mo}}, \bibinfo
  {author} {\bibfnamefont {Z.}~\bibnamefont {Hussain}}, \bibinfo {author}
  {\bibfnamefont {T.~P.}\ \bibnamefont {Devereaux}}, \bibinfo {author}
  {\bibfnamefont {Z.-X.}\ \bibnamefont {Shen}}, \ and\ \bibinfo {author}
  {\bibfnamefont {D.~H.}\ \bibnamefont {Lu}},\ }\href@noop {} {\bibfield
  {journal} {\bibinfo  {journal} {arXiv 1503.01556}\ } (\bibinfo {year}
  {2015}{\natexlab{b}})}\BibitemShut {NoStop}%
\bibitem [{\citenamefont {Suzuki}\ \emph {et~al.}(2015)\citenamefont {Suzuki},
  \citenamefont {Shimojima}, \citenamefont {Sonobe}, \citenamefont {Nakamura},
  \citenamefont {Sakano}, \citenamefont {Tsuji}, \citenamefont {Omachi},
  \citenamefont {Yoshioka}, \citenamefont {Kuwata-Gonokami}, \citenamefont
  {Watashige}, \citenamefont {Kobayashi}, \citenamefont {Kasahara},
  \citenamefont {Shibauchi}, \citenamefont {Matsuda}, \citenamefont {Yamakawa},
  \citenamefont {Kontani},\ and\ \citenamefont {Ishizaka}}]{Suzuki2015}%
  \BibitemOpen
  \bibfield  {author} {\bibinfo {author} {\bibfnamefont {Y.}~\bibnamefont
  {Suzuki}}, \bibinfo {author} {\bibfnamefont {T.}~\bibnamefont {Shimojima}},
  \bibinfo {author} {\bibfnamefont {T.}~\bibnamefont {Sonobe}}, \bibinfo
  {author} {\bibfnamefont {A.}~\bibnamefont {Nakamura}}, \bibinfo {author}
  {\bibfnamefont {M.}~\bibnamefont {Sakano}}, \bibinfo {author} {\bibfnamefont
  {H.}~\bibnamefont {Tsuji}}, \bibinfo {author} {\bibfnamefont
  {J.}~\bibnamefont {Omachi}}, \bibinfo {author} {\bibfnamefont
  {K.}~\bibnamefont {Yoshioka}}, \bibinfo {author} {\bibfnamefont
  {M.}~\bibnamefont {Kuwata-Gonokami}}, \bibinfo {author} {\bibfnamefont
  {T.}~\bibnamefont {Watashige}}, \bibinfo {author} {\bibfnamefont
  {R.}~\bibnamefont {Kobayashi}}, \bibinfo {author} {\bibfnamefont
  {S.}~\bibnamefont {Kasahara}}, \bibinfo {author} {\bibfnamefont
  {T.}~\bibnamefont {Shibauchi}}, \bibinfo {author} {\bibfnamefont
  {Y.}~\bibnamefont {Matsuda}}, \bibinfo {author} {\bibfnamefont
  {Y.}~\bibnamefont {Yamakawa}}, \bibinfo {author} {\bibfnamefont
  {H.}~\bibnamefont {Kontani}}, \ and\ \bibinfo {author} {\bibfnamefont
  {K.}~\bibnamefont {Ishizaka}},\ }\href@noop {} {\  (\bibinfo {year}
  {2015})},\ \Eprint {http://arxiv.org/abs/1504.00980} {arXiv:1504.00980}
  \BibitemShut {NoStop}%
\bibitem [{\citenamefont {Terashima}\ \emph {et~al.}(2015)\citenamefont
  {Terashima}, \citenamefont {Kikugawa}, \citenamefont {Kasahara},
  \citenamefont {Watashige}, \citenamefont {Shibauchi}, \citenamefont
  {Matsuda}, \citenamefont {Wolf}, \citenamefont {B\"{o}hmer}, \citenamefont
  {Hardy}, \citenamefont {Meingast}, \citenamefont {L\"{o}hneysen},\ and\
  \citenamefont {Uji}}]{Terashima2015}%
  \BibitemOpen
  \bibfield  {author} {\bibinfo {author} {\bibfnamefont {T.}~\bibnamefont
  {Terashima}}, \bibinfo {author} {\bibfnamefont {N.}~\bibnamefont {Kikugawa}},
  \bibinfo {author} {\bibfnamefont {S.}~\bibnamefont {Kasahara}}, \bibinfo
  {author} {\bibfnamefont {T.}~\bibnamefont {Watashige}}, \bibinfo {author}
  {\bibfnamefont {T.}~\bibnamefont {Shibauchi}}, \bibinfo {author}
  {\bibfnamefont {Y.}~\bibnamefont {Matsuda}}, \bibinfo {author} {\bibfnamefont
  {T.}~\bibnamefont {Wolf}}, \bibinfo {author} {\bibfnamefont {A.~E.}\
  \bibnamefont {B\"{o}hmer}}, \bibinfo {author} {\bibfnamefont
  {F.}~\bibnamefont {Hardy}}, \bibinfo {author} {\bibfnamefont
  {C.}~\bibnamefont {Meingast}}, \bibinfo {author} {\bibfnamefont {H.~v.}\
  \bibnamefont {L\"{o}hneysen}}, \ and\ \bibinfo {author} {\bibfnamefont
  {S.}~\bibnamefont {Uji}},\ }\href {\doibase 10.7566/JPSJ.84.063701}
  {\bibfield  {journal} {\bibinfo  {journal} {J. Phys. Soc. Japan}\ }\textbf
  {\bibinfo {volume} {84}},\ \bibinfo {pages} {063701} (\bibinfo {year}
  {2015})}\BibitemShut {NoStop}%
\bibitem [{\citenamefont {Kn\"{o}ner}\ \emph {et~al.}(2015)\citenamefont
  {Kn\"{o}ner}, \citenamefont {Zielke}, \citenamefont {K\"{o}hler},
  \citenamefont {Wolf}, \citenamefont {Wolf}, \citenamefont {Wang},
  \citenamefont {B\"{o}hmer}, \citenamefont {Meingast},\ and\ \citenamefont
  {Lang}}]{Knoner2015}%
  \BibitemOpen
  \bibfield  {author} {\bibinfo {author} {\bibfnamefont {S.}~\bibnamefont
  {Kn\"{o}ner}}, \bibinfo {author} {\bibfnamefont {D.}~\bibnamefont {Zielke}},
  \bibinfo {author} {\bibfnamefont {S.}~\bibnamefont {K\"{o}hler}}, \bibinfo
  {author} {\bibfnamefont {B.}~\bibnamefont {Wolf}}, \bibinfo {author}
  {\bibfnamefont {T.}~\bibnamefont {Wolf}}, \bibinfo {author} {\bibfnamefont
  {L.}~\bibnamefont {Wang}}, \bibinfo {author} {\bibfnamefont {A.}~\bibnamefont
  {B\"{o}hmer}}, \bibinfo {author} {\bibfnamefont {C.}~\bibnamefont
  {Meingast}}, \ and\ \bibinfo {author} {\bibfnamefont {M.}~\bibnamefont
  {Lang}},\ }\href {\doibase 10.1103/PhysRevB.91.174510} {\bibfield  {journal}
  {\bibinfo  {journal} {Phys. Rev. B}\ }\textbf {\bibinfo {volume} {91}},\
  \bibinfo {pages} {174510} (\bibinfo {year} {2015})}\BibitemShut {NoStop}%
\bibitem [{\citenamefont {Tan}\ \emph {et~al.}(2013)\citenamefont {Tan},
  \citenamefont {Zhang}, \citenamefont {Xia}, \citenamefont {Ye}, \citenamefont
  {Chen}, \citenamefont {Xie}, \citenamefont {Peng}, \citenamefont {Xu},
  \citenamefont {Fan}, \citenamefont {Xu}, \citenamefont {Jiang}, \citenamefont
  {Zhang}, \citenamefont {Lai}, \citenamefont {Xiang}, \citenamefont {Hu},
  \citenamefont {Xie},\ and\ \citenamefont {Feng}}]{Tan2013}%
  \BibitemOpen
  \bibfield  {author} {\bibinfo {author} {\bibfnamefont {S.}~\bibnamefont
  {Tan}}, \bibinfo {author} {\bibfnamefont {Y.}~\bibnamefont {Zhang}}, \bibinfo
  {author} {\bibfnamefont {M.}~\bibnamefont {Xia}}, \bibinfo {author}
  {\bibfnamefont {Z.}~\bibnamefont {Ye}}, \bibinfo {author} {\bibfnamefont
  {F.}~\bibnamefont {Chen}}, \bibinfo {author} {\bibfnamefont {X.}~\bibnamefont
  {Xie}}, \bibinfo {author} {\bibfnamefont {R.}~\bibnamefont {Peng}}, \bibinfo
  {author} {\bibfnamefont {D.}~\bibnamefont {Xu}}, \bibinfo {author}
  {\bibfnamefont {Q.}~\bibnamefont {Fan}}, \bibinfo {author} {\bibfnamefont
  {H.}~\bibnamefont {Xu}}, \bibinfo {author} {\bibfnamefont {J.}~\bibnamefont
  {Jiang}}, \bibinfo {author} {\bibfnamefont {T.}~\bibnamefont {Zhang}},
  \bibinfo {author} {\bibfnamefont {X.}~\bibnamefont {Lai}}, \bibinfo {author}
  {\bibfnamefont {T.}~\bibnamefont {Xiang}}, \bibinfo {author} {\bibfnamefont
  {J.}~\bibnamefont {Hu}}, \bibinfo {author} {\bibfnamefont {B.}~\bibnamefont
  {Xie}}, \ and\ \bibinfo {author} {\bibfnamefont {D.}~\bibnamefont {Feng}},\
  }\href {\doibase 10.1038/nmat3654} {\bibfield  {journal} {\bibinfo  {journal}
  {Nat. Mater.}\ }\textbf {\bibinfo {volume} {12}},\ \bibinfo {pages} {634}
  (\bibinfo {year} {2013})}\BibitemShut {NoStop}%
\bibitem [{\citenamefont {Maletz}\ \emph {et~al.}(2014)\citenamefont {Maletz},
  \citenamefont {Zabolotnyy}, \citenamefont {Evtushinsky}, \citenamefont
  {Thirupathaiah}, \citenamefont {Wolter}, \citenamefont {Harnagea},
  \citenamefont {Yaresko}, \citenamefont {Vasiliev}, \citenamefont {Chareev},
  \citenamefont {B\"{o}hmer}, \citenamefont {Hardy}, \citenamefont {Wolf},
  \citenamefont {Meingast}, \citenamefont {Rienks}, \citenamefont
  {B\"{u}chner},\ and\ \citenamefont {Borisenko}}]{Maletz2014}%
  \BibitemOpen
  \bibfield  {author} {\bibinfo {author} {\bibfnamefont {J.}~\bibnamefont
  {Maletz}}, \bibinfo {author} {\bibfnamefont {V.~B.}\ \bibnamefont
  {Zabolotnyy}}, \bibinfo {author} {\bibfnamefont {D.~V.}\ \bibnamefont
  {Evtushinsky}}, \bibinfo {author} {\bibfnamefont {S.}~\bibnamefont
  {Thirupathaiah}}, \bibinfo {author} {\bibfnamefont {A.~U.~B.}\ \bibnamefont
  {Wolter}}, \bibinfo {author} {\bibfnamefont {L.}~\bibnamefont {Harnagea}},
  \bibinfo {author} {\bibfnamefont {A.~N.}\ \bibnamefont {Yaresko}}, \bibinfo
  {author} {\bibfnamefont {A.~N.}\ \bibnamefont {Vasiliev}}, \bibinfo {author}
  {\bibfnamefont {D.~A.}\ \bibnamefont {Chareev}}, \bibinfo {author}
  {\bibfnamefont {A.~E.}\ \bibnamefont {B\"{o}hmer}}, \bibinfo {author}
  {\bibfnamefont {F.}~\bibnamefont {Hardy}}, \bibinfo {author} {\bibfnamefont
  {T.}~\bibnamefont {Wolf}}, \bibinfo {author} {\bibfnamefont {C.}~\bibnamefont
  {Meingast}}, \bibinfo {author} {\bibfnamefont {E.~D.~L.}\ \bibnamefont
  {Rienks}}, \bibinfo {author} {\bibfnamefont {B.}~\bibnamefont {B\"{u}chner}},
  \ and\ \bibinfo {author} {\bibfnamefont {S.~V.}\ \bibnamefont {Borisenko}},\
  }\href {\doibase 10.1103/PhysRevB.89.220506} {\bibfield  {journal} {\bibinfo
  {journal} {Phys. Rev. B}\ }\textbf {\bibinfo {volume} {89}},\ \bibinfo
  {pages} {220506} (\bibinfo {year} {2014})}\BibitemShut {NoStop}%
\bibitem [{\citenamefont {Chareev}\ \emph {et~al.}(2013)\citenamefont
  {Chareev}, \citenamefont {Osadchii}, \citenamefont {Kuzmicheva},
  \citenamefont {Lin}, \citenamefont {Kuzmichev}, \citenamefont {Volkova},\
  and\ \citenamefont {Vasiliev}}]{Chareev2013}%
  \BibitemOpen
  \bibfield  {author} {\bibinfo {author} {\bibfnamefont {D.}~\bibnamefont
  {Chareev}}, \bibinfo {author} {\bibfnamefont {E.}~\bibnamefont {Osadchii}},
  \bibinfo {author} {\bibfnamefont {T.}~\bibnamefont {Kuzmicheva}}, \bibinfo
  {author} {\bibfnamefont {J.-Y.}\ \bibnamefont {Lin}}, \bibinfo {author}
  {\bibfnamefont {S.}~\bibnamefont {Kuzmichev}}, \bibinfo {author}
  {\bibfnamefont {O.}~\bibnamefont {Volkova}}, \ and\ \bibinfo {author}
  {\bibfnamefont {A.}~\bibnamefont {Vasiliev}},\ }\href {\doibase
  10.1039/c2ce26857d} {\bibfield  {journal} {\bibinfo  {journal}
  {CrystEngComm}\ }\textbf {\bibinfo {volume} {15}},\ \bibinfo {pages} {1989}
  (\bibinfo {year} {2013})}\BibitemShut {NoStop}%
\bibitem [{\citenamefont {B\"{o}hmer}\ \emph {et~al.}(2013)\citenamefont
  {B\"{o}hmer}, \citenamefont {Hardy}, \citenamefont {Eilers}, \citenamefont
  {Ernst}, \citenamefont {Adelmann}, \citenamefont {Schweiss}, \citenamefont
  {Wolf},\ and\ \citenamefont {Meingast}}]{Bohmer2013}%
  \BibitemOpen
  \bibfield  {author} {\bibinfo {author} {\bibfnamefont {A.~E.}\ \bibnamefont
  {B\"{o}hmer}}, \bibinfo {author} {\bibfnamefont {F.}~\bibnamefont {Hardy}},
  \bibinfo {author} {\bibfnamefont {F.}~\bibnamefont {Eilers}}, \bibinfo
  {author} {\bibfnamefont {D.}~\bibnamefont {Ernst}}, \bibinfo {author}
  {\bibfnamefont {P.}~\bibnamefont {Adelmann}}, \bibinfo {author}
  {\bibfnamefont {P.}~\bibnamefont {Schweiss}}, \bibinfo {author}
  {\bibfnamefont {T.}~\bibnamefont {Wolf}}, \ and\ \bibinfo {author}
  {\bibfnamefont {C.}~\bibnamefont {Meingast}},\ }\href {\doibase
  10.1103/PhysRevB.87.180505} {\bibfield  {journal} {\bibinfo  {journal} {Phys.
  Rev. B}\ }\textbf {\bibinfo {volume} {87}},\ \bibinfo {pages} {180505}
  (\bibinfo {year} {2013})}\BibitemShut {NoStop}%
\bibitem [{\citenamefont {Blaha}\ \emph {et~al.}(2001)\citenamefont {Blaha},
  \citenamefont {Schwarz}, \citenamefont {Madsen}, \citenamefont {Kvasnicka},\
  and\ \citenamefont {Luitz}}]{Wien2k}%
  \BibitemOpen
  \bibfield  {author} {\bibinfo {author} {\bibfnamefont {P.}~\bibnamefont
  {Blaha}}, \bibinfo {author} {\bibfnamefont {K.}~\bibnamefont {Schwarz}},
  \bibinfo {author} {\bibfnamefont {G.}~\bibnamefont {Madsen}}, \bibinfo
  {author} {\bibfnamefont {D.}~\bibnamefont {Kvasnicka}}, \ and\ \bibinfo
  {author} {\bibfnamefont {J.}~\bibnamefont {Luitz}},\ }\href@noop {} {\emph
  {\bibinfo {title} {WIEN2k}}}\ (\bibinfo  {publisher} {Techn. Universit\"{a}t
  Wien, Austria},\ \bibinfo {year} {2001})\BibitemShut {NoStop}%
\bibitem [{Note1()}]{Note1}%
  \BibitemOpen
  \bibinfo {note} {Since FeSe does not show magnetic order ARPES spectra are
  compared to non-spin-polarized band structure, with the relaxed lattice
  parameters $a = 3.7651$ \r A, $c$ = 5.5178 \r A, $z_{Se}$ = 0.24128 from
  Ref.~\protect \rev@citealp {Kumar2012}}\BibitemShut {NoStop}%
\bibitem [{\citenamefont {Mizuguchi}\ \emph {et~al.}(2009)\citenamefont
  {Mizuguchi}, \citenamefont {Tomioka}, \citenamefont {Tsuda}, \citenamefont
  {Yamaguchi},\ and\ \citenamefont {Takano}}]{Mizuguchi2009}%
  \BibitemOpen
  \bibfield  {author} {\bibinfo {author} {\bibfnamefont {Y.}~\bibnamefont
  {Mizuguchi}}, \bibinfo {author} {\bibfnamefont {F.}~\bibnamefont {Tomioka}},
  \bibinfo {author} {\bibfnamefont {S.}~\bibnamefont {Tsuda}}, \bibinfo
  {author} {\bibfnamefont {T.}~\bibnamefont {Yamaguchi}}, \ and\ \bibinfo
  {author} {\bibfnamefont {Y.}~\bibnamefont {Takano}},\ }\href {\doibase
  10.1143/JPSJ.78.074712} {\bibfield  {journal} {\bibinfo  {journal} {J. Phys.
  Soc. Japan}\ }\textbf {\bibinfo {volume} {78}},\ \bibinfo {pages} {074712}
  (\bibinfo {year} {2009})}\BibitemShut {NoStop}%
\bibitem [{Note2()}]{Note2}%
  \BibitemOpen
  \bibinfo {note} {We have never observed the outer $d_{xy}$ electron band,
  even in the second Brillouin zone where the weak signal due to matrix
  elements effects might be expected to be larger}\BibitemShut {NoStop}%
\bibitem [{\citenamefont {Coldea}(2015)}]{FeSeS_QOs}%
  \BibitemOpen
  \bibfield  {author} {\bibinfo {author} {\bibfnamefont {A.}~\bibnamefont
  {Coldea}},\ }\href@noop {} {\bibfield  {journal} {\bibinfo  {journal} {in
  preparation}\ } (\bibinfo {year} {2015})}\BibitemShut {NoStop}%
\bibitem [{\citenamefont {Fernandes}\ and\ \citenamefont
  {Vafek}(2014)}]{Fernandes2014b}%
  \BibitemOpen
  \bibfield  {author} {\bibinfo {author} {\bibfnamefont {R.~M.}\ \bibnamefont
  {Fernandes}}\ and\ \bibinfo {author} {\bibfnamefont {O.}~\bibnamefont
  {Vafek}},\ }\href {\doibase 10.1103/PhysRevB.90.214514} {\bibfield  {journal}
  {\bibinfo  {journal} {Phys. Rev. B}\ }\textbf {\bibinfo {volume} {90}},\
  \bibinfo {pages} {214514} (\bibinfo {year} {2014})}\BibitemShut {NoStop}%
\bibitem [{\citenamefont {Watson}\ \emph {et~al.}()\citenamefont {Watson},
  \citenamefont {Yamashita}, \citenamefont {Kasahara}, \citenamefont {Knafo},
  \citenamefont {Nardone}, \citenamefont {B\'{e}ard}, \citenamefont {Hardy},
  \citenamefont {McCollam}, \citenamefont {Narayanan}, \citenamefont {Blake},
  \citenamefont {Wolf}, \citenamefont {Haghighirad}, \citenamefont {Meingast},
  \citenamefont {Schofield}, \citenamefont {von L\"{o}hneysen}, \citenamefont
  {Matsuda}, \citenamefont {Coldea},\ and\ \citenamefont
  {Shibauchi}}]{Watson2015b}%
  \BibitemOpen
  \bibfield  {author} {\bibinfo {author} {\bibfnamefont {M.~D.}\ \bibnamefont
  {Watson}}, \bibinfo {author} {\bibfnamefont {T.}~\bibnamefont {Yamashita}},
  \bibinfo {author} {\bibfnamefont {S.}~\bibnamefont {Kasahara}}, \bibinfo
  {author} {\bibfnamefont {W.}~\bibnamefont {Knafo}}, \bibinfo {author}
  {\bibfnamefont {M.}~\bibnamefont {Nardone}}, \bibinfo {author} {\bibfnamefont
  {J.}~\bibnamefont {B\'{e}ard}}, \bibinfo {author} {\bibfnamefont
  {F.}~\bibnamefont {Hardy}}, \bibinfo {author} {\bibfnamefont
  {A.}~\bibnamefont {McCollam}}, \bibinfo {author} {\bibfnamefont
  {A.}~\bibnamefont {Narayanan}}, \bibinfo {author} {\bibfnamefont {S.~F.}\
  \bibnamefont {Blake}}, \bibinfo {author} {\bibfnamefont {T.}~\bibnamefont
  {Wolf}}, \bibinfo {author} {\bibfnamefont {A.~A.}\ \bibnamefont
  {Haghighirad}}, \bibinfo {author} {\bibfnamefont {C.}~\bibnamefont
  {Meingast}}, \bibinfo {author} {\bibfnamefont {A.~J.}\ \bibnamefont
  {Schofield}}, \bibinfo {author} {\bibfnamefont {H.}~\bibnamefont {von
  L\"{o}hneysen}}, \bibinfo {author} {\bibfnamefont {Y.}~\bibnamefont
  {Matsuda}}, \bibinfo {author} {\bibfnamefont {A.~I.}\ \bibnamefont {Coldea}},
  \ and\ \bibinfo {author} {\bibfnamefont {T.}~\bibnamefont {Shibauchi}},\
  }\href@noop {} {\bibfield  {journal} {\bibinfo  {journal} {arXiv:
  1502.02922v1}\ }}\Eprint {http://arxiv.org/abs/1502.02922v1}
  {arXiv:1502.02922v1} \BibitemShut {NoStop}%
\bibitem [{\citenamefont {Terashima}\ \emph {et~al.}(2014)\citenamefont
  {Terashima}, \citenamefont {Kikugawa}, \citenamefont {Kiswandhi},
  \citenamefont {Choi}, \citenamefont {Brooks}, \citenamefont {Kasahara},
  \citenamefont {Watashige}, \citenamefont {Ikeda}, \citenamefont {Shibauchi},
  \citenamefont {Matsuda}, \citenamefont {Wolf}, \citenamefont {B\"{o}hmer},
  \citenamefont {Hardy}, \citenamefont {Meingast}, \citenamefont
  {L\"{o}hneysen}, \citenamefont {Suzuki}, \citenamefont {Arita},\ and\
  \citenamefont {Uji}}]{Terashima2014}%
  \BibitemOpen
  \bibfield  {author} {\bibinfo {author} {\bibfnamefont {T.}~\bibnamefont
  {Terashima}}, \bibinfo {author} {\bibfnamefont {N.}~\bibnamefont {Kikugawa}},
  \bibinfo {author} {\bibfnamefont {A.}~\bibnamefont {Kiswandhi}}, \bibinfo
  {author} {\bibfnamefont {E.-S.}\ \bibnamefont {Choi}}, \bibinfo {author}
  {\bibfnamefont {J.~S.}\ \bibnamefont {Brooks}}, \bibinfo {author}
  {\bibfnamefont {S.}~\bibnamefont {Kasahara}}, \bibinfo {author}
  {\bibfnamefont {T.}~\bibnamefont {Watashige}}, \bibinfo {author}
  {\bibfnamefont {H.}~\bibnamefont {Ikeda}}, \bibinfo {author} {\bibfnamefont
  {T.}~\bibnamefont {Shibauchi}}, \bibinfo {author} {\bibfnamefont
  {Y.}~\bibnamefont {Matsuda}}, \bibinfo {author} {\bibfnamefont
  {T.}~\bibnamefont {Wolf}}, \bibinfo {author} {\bibfnamefont {A.~E.}\
  \bibnamefont {B\"{o}hmer}}, \bibinfo {author} {\bibfnamefont
  {F.}~\bibnamefont {Hardy}}, \bibinfo {author} {\bibfnamefont
  {C.}~\bibnamefont {Meingast}}, \bibinfo {author} {\bibfnamefont {H.~V.}\
  \bibnamefont {L\"{o}hneysen}}, \bibinfo {author} {\bibfnamefont {M.-T.}\
  \bibnamefont {Suzuki}}, \bibinfo {author} {\bibfnamefont {R.}~\bibnamefont
  {Arita}}, \ and\ \bibinfo {author} {\bibfnamefont {S.}~\bibnamefont {Uji}},\
  }\href {\doibase 10.1103/PhysRevB.90.144517} {\bibfield  {journal} {\bibinfo
  {journal} {Phys. Rev. B}\ }\textbf {\bibinfo {volume} {90}},\ \bibinfo
  {pages} {155517} (\bibinfo {year} {2014})}\BibitemShut {NoStop}%
\bibitem [{\citenamefont {Abdel-Hafiez}\ \emph {et~al.}(2015)\citenamefont
  {Abdel-Hafiez}, \citenamefont {Zhang}, \citenamefont {Cao}, \citenamefont
  {Duan}, \citenamefont {Karapetrov}, \citenamefont {Pudalov}, \citenamefont
  {Vlasenko}, \citenamefont {Sadakov}, \citenamefont {Knyazev}, \citenamefont
  {Romanova}, \citenamefont {Chareev}, \citenamefont {Volkova}, \citenamefont
  {Vasiliev},\ and\ \citenamefont {Chen}}]{Abdel-Hafiez2015}%
  \BibitemOpen
  \bibfield  {author} {\bibinfo {author} {\bibfnamefont {M.}~\bibnamefont
  {Abdel-Hafiez}}, \bibinfo {author} {\bibfnamefont {Y.-Y.}\ \bibnamefont
  {Zhang}}, \bibinfo {author} {\bibfnamefont {Z.-Y.}\ \bibnamefont {Cao}},
  \bibinfo {author} {\bibfnamefont {C.-G.}\ \bibnamefont {Duan}}, \bibinfo
  {author} {\bibfnamefont {G.}~\bibnamefont {Karapetrov}}, \bibinfo {author}
  {\bibfnamefont {V.~M.}\ \bibnamefont {Pudalov}}, \bibinfo {author}
  {\bibfnamefont {V.~A.}\ \bibnamefont {Vlasenko}}, \bibinfo {author}
  {\bibfnamefont {A.~V.}\ \bibnamefont {Sadakov}}, \bibinfo {author}
  {\bibfnamefont {D.~A.}\ \bibnamefont {Knyazev}}, \bibinfo {author}
  {\bibfnamefont {T.~A.}\ \bibnamefont {Romanova}}, \bibinfo {author}
  {\bibfnamefont {D.~A.}\ \bibnamefont {Chareev}}, \bibinfo {author}
  {\bibfnamefont {O.~S.}\ \bibnamefont {Volkova}}, \bibinfo {author}
  {\bibfnamefont {A.~N.}\ \bibnamefont {Vasiliev}}, \ and\ \bibinfo {author}
  {\bibfnamefont {X.-J.}\ \bibnamefont {Chen}},\ }\href {\doibase
  10.1103/PhysRevB.91.165109} {\bibfield  {journal} {\bibinfo  {journal} {Phys.
  Rev. B}\ }\textbf {\bibinfo {volume} {91}},\ \bibinfo {pages} {165109}
  (\bibinfo {year} {2015})}\BibitemShut {NoStop}%
\bibitem [{\citenamefont {Bendele}\ \emph {et~al.}(2010)\citenamefont
  {Bendele}, \citenamefont {Amato}, \citenamefont {Conder}, \citenamefont
  {Elender}, \citenamefont {Keller}, \citenamefont {Klauss}, \citenamefont
  {Luetkens}, \citenamefont {Pomjakushina}, \citenamefont {Raselli},\ and\
  \citenamefont {Khasanov}}]{Bendele2010}%
  \BibitemOpen
  \bibfield  {author} {\bibinfo {author} {\bibfnamefont {M.}~\bibnamefont
  {Bendele}}, \bibinfo {author} {\bibfnamefont {A.}~\bibnamefont {Amato}},
  \bibinfo {author} {\bibfnamefont {K.}~\bibnamefont {Conder}}, \bibinfo
  {author} {\bibfnamefont {M.}~\bibnamefont {Elender}}, \bibinfo {author}
  {\bibfnamefont {H.}~\bibnamefont {Keller}}, \bibinfo {author} {\bibfnamefont
  {H.-H.}\ \bibnamefont {Klauss}}, \bibinfo {author} {\bibfnamefont
  {H.}~\bibnamefont {Luetkens}}, \bibinfo {author} {\bibfnamefont
  {E.}~\bibnamefont {Pomjakushina}}, \bibinfo {author} {\bibfnamefont
  {A.}~\bibnamefont {Raselli}}, \ and\ \bibinfo {author} {\bibfnamefont
  {R.}~\bibnamefont {Khasanov}},\ }\href {\doibase
  10.1103/PhysRevLett.104.087003} {\bibfield  {journal} {\bibinfo  {journal}
  {Phys. Rev. Lett.}\ }\textbf {\bibinfo {volume} {104}},\ \bibinfo {pages}
  {087003} (\bibinfo {year} {2010})}\BibitemShut {NoStop}%
\bibitem [{\citenamefont {Medvedev}\ \emph {et~al.}(2009)\citenamefont
  {Medvedev}, \citenamefont {McQueen}, \citenamefont {Troyan}, \citenamefont
  {Palasyuk}, \citenamefont {Eremets}, \citenamefont {Cava}, \citenamefont
  {Naghavi}, \citenamefont {Casper}, \citenamefont {Ksenofontov}, \citenamefont
  {Wortmann},\ and\ \citenamefont {Felser}}]{Medvedev2009}%
  \BibitemOpen
  \bibfield  {author} {\bibinfo {author} {\bibfnamefont {S.}~\bibnamefont
  {Medvedev}}, \bibinfo {author} {\bibfnamefont {T.~M.}\ \bibnamefont
  {McQueen}}, \bibinfo {author} {\bibfnamefont {I.~A.}\ \bibnamefont {Troyan}},
  \bibinfo {author} {\bibfnamefont {T.}~\bibnamefont {Palasyuk}}, \bibinfo
  {author} {\bibfnamefont {M.~I.}\ \bibnamefont {Eremets}}, \bibinfo {author}
  {\bibfnamefont {R.~J.}\ \bibnamefont {Cava}}, \bibinfo {author}
  {\bibfnamefont {S.}~\bibnamefont {Naghavi}}, \bibinfo {author} {\bibfnamefont
  {F.}~\bibnamefont {Casper}}, \bibinfo {author} {\bibfnamefont
  {V.}~\bibnamefont {Ksenofontov}}, \bibinfo {author} {\bibfnamefont
  {G.}~\bibnamefont {Wortmann}}, \ and\ \bibinfo {author} {\bibfnamefont
  {C.}~\bibnamefont {Felser}},\ }\href {\doibase 10.1038/nmat2491} {\bibfield
  {journal} {\bibinfo  {journal} {Nat. Mater.}\ }\textbf {\bibinfo {volume}
  {8}},\ \bibinfo {pages} {630} (\bibinfo {year} {2009})}\BibitemShut {NoStop}%
\bibitem [{\citenamefont {Tomita}\ \emph {et~al.}(2015)\citenamefont {Tomita},
  \citenamefont {Takahashi}, \citenamefont {Takahashi}, \citenamefont {Okada},
  \citenamefont {Mizuguchi}, \citenamefont {Takano}, \citenamefont {Nakano},
  \citenamefont {Matsubayashi},\ and\ \citenamefont {Uwatoko}}]{Tomita2015}%
  \BibitemOpen
  \bibfield  {author} {\bibinfo {author} {\bibfnamefont {T.}~\bibnamefont
  {Tomita}}, \bibinfo {author} {\bibfnamefont {H.}~\bibnamefont {Takahashi}},
  \bibinfo {author} {\bibfnamefont {H.}~\bibnamefont {Takahashi}}, \bibinfo
  {author} {\bibfnamefont {H.}~\bibnamefont {Okada}}, \bibinfo {author}
  {\bibfnamefont {Y.}~\bibnamefont {Mizuguchi}}, \bibinfo {author}
  {\bibfnamefont {Y.}~\bibnamefont {Takano}}, \bibinfo {author} {\bibfnamefont
  {S.}~\bibnamefont {Nakano}}, \bibinfo {author} {\bibfnamefont
  {K.}~\bibnamefont {Matsubayashi}}, \ and\ \bibinfo {author} {\bibfnamefont
  {Y.}~\bibnamefont {Uwatoko}},\ }\href {\doibase 10.7566/JPSJ.84.024713}
  {\bibfield  {journal} {\bibinfo  {journal} {J. Phys. Soc. Japan}\ }\textbf
  {\bibinfo {volume} {84}},\ \bibinfo {pages} {24713} (\bibinfo {year}
  {2015})}\BibitemShut {NoStop}%
\bibitem [{\citenamefont {Imai}\ \emph {et~al.}(2009)\citenamefont {Imai},
  \citenamefont {Ahilan}, \citenamefont {Ning}, \citenamefont {McQueen},\ and\
  \citenamefont {Cava}}]{Imai2009}%
  \BibitemOpen
  \bibfield  {author} {\bibinfo {author} {\bibfnamefont {T.}~\bibnamefont
  {Imai}}, \bibinfo {author} {\bibfnamefont {K.}~\bibnamefont {Ahilan}},
  \bibinfo {author} {\bibfnamefont {F.~L.}\ \bibnamefont {Ning}}, \bibinfo
  {author} {\bibfnamefont {T.~M.}\ \bibnamefont {McQueen}}, \ and\ \bibinfo
  {author} {\bibfnamefont {R.~J.}\ \bibnamefont {Cava}},\ }\href {\doibase
  10.1103/PhysRevLett.102.177005} {\bibfield  {journal} {\bibinfo  {journal}
  {Phys. Rev. Lett.}\ }\textbf {\bibinfo {volume} {102}},\ \bibinfo {pages}
  {177005} (\bibinfo {year} {2009})}\BibitemShut {NoStop}%
\bibitem [{\citenamefont {Jiang}\ \emph {et~al.}(2015)\citenamefont {Jiang},
  \citenamefont {Hu}, \citenamefont {Ding},\ and\ \citenamefont
  {Wang}}]{Jiang2015}%
  \BibitemOpen
  \bibfield  {author} {\bibinfo {author} {\bibfnamefont {K.}~\bibnamefont
  {Jiang}}, \bibinfo {author} {\bibfnamefont {J.}~\bibnamefont {Hu}}, \bibinfo
  {author} {\bibfnamefont {H.}~\bibnamefont {Ding}}, \ and\ \bibinfo {author}
  {\bibfnamefont {Z.}~\bibnamefont {Wang}},\ }\href
  {http://arxiv.org/abs/1508.00588} {\  (\bibinfo {year} {2015})},\ \Eprint
  {http://arxiv.org/abs/1508.00588} {arXiv:1508.00588} \BibitemShut {NoStop}%
\bibitem [{\citenamefont {Chubukov}(2015)}]{ChubukovPC}%
  \BibitemOpen
  \bibfield  {author} {\bibinfo {author} {\bibfnamefont {A.}~\bibnamefont
  {Chubukov}},\ }\href@noop {} {\bibfield  {journal} {\bibinfo  {journal}
  {private communication}\ } (\bibinfo {year} {2015})}\BibitemShut {NoStop}%
\bibitem [{\citenamefont {Kumar}\ \emph {et~al.}(2012)\citenamefont {Kumar},
  \citenamefont {Auluck}, \citenamefont {Ahluwalia},\ and\ \citenamefont
  {Awana}}]{Kumar2012}%
  \BibitemOpen
  \bibfield  {author} {\bibinfo {author} {\bibfnamefont {J.}~\bibnamefont
  {Kumar}}, \bibinfo {author} {\bibfnamefont {S.}~\bibnamefont {Auluck}},
  \bibinfo {author} {\bibfnamefont {P.~K.}\ \bibnamefont {Ahluwalia}}, \ and\
  \bibinfo {author} {\bibfnamefont {V.~P.~S.}\ \bibnamefont {Awana}},\ }\href
  {\doibase 10.1088/0953-2048/25/9/095002} {\bibfield  {journal} {\bibinfo
  {journal} {Supercond. Sci. Technol.}\ }\textbf {\bibinfo {volume} {25}},\
  \bibinfo {pages} {095002} (\bibinfo {year} {2012})}\BibitemShut {NoStop}%
\end{thebibliography}%

%

\end{document}